\documentclass[aps,prl,twocolumn,groupedaddress,showpacs,superscriptaddress,scrartcl,longbibliography]{revtex4-2}
\usepackage{xcolor,comment,amssymb,amsmath,graphicx,epstopdf,hyperref,amsthm,dsfont,enumerate,bm,color}

\begin{document}
\interfootnotelinepenalty=10000

\title{Unconditionally secure relativistic multi-party biased coin flipping and die rolling}

\author{Dami\'an Pital\'ua-Garc\'ia}
\email{D.Pitalua-Garcia@damtp.cam.ac.uk}
\affiliation{Centre for Quantum Information and Foundations, DAMTP, Centre for Mathematical Sciences, University of Cambridge, Wilberforce Road, Cambridge, CB3 0WA, U.K.}

\date{\today}

\begin{abstract}
We introduce relativistic multi-party biased die rolling protocols, generalizing coin
flipping to $M \geq 2$ parties and to $N \geq 2$ outcomes for any chosen outcome biases, and show them unconditionally secure. Our results prove that the most general random secure multi-party computation, where all parties receive the output and there is no secret input by any party, can be implemented with unconditional security. Our protocols extend Kent’s [A. Kent, Phys. Rev. Lett. \textbf{83}, 5382 (1999)] two-party unbiased coin flipping protocol, do not require any quantum communication, are practical to implement with current technology, and to  our knowledge are the first multi-party relativistic cryptographic protocols.




\end{abstract}


\maketitle

\section{Introduction}

$M$ mistrustful parties at different locations roll a $N-$faced die via some agreed protocol $\mathcal{R}$ in such a way that if the $k$th party follows $\mathcal{R}$ honestly and the other parties deviate arbitrarily from $\mathcal{R}$ then the outcome $o$ is obtained with a probability $P(o)$ satisfying $\lvert P(o)-P_o\rvert\leq\delta$, 
for all $o\in\mathbb{Z}_N=\{0,1,\ldots,N-1\}$, for all $k\in[M]=\{1,2,\ldots,M\}$, for agreed integers $M,N\geq 2$ and for an agreed probability distribution $\mathcal{P}=\{P_o\}_{o=0}^{N-1}$. This task is called \emph{$M-$party biased $N-$faced die rolling}, or simply \emph{die rolling}, and is the most general type 
of random secure multi-party computation where all parties receive the output of the computation and there is no secret input by any party \cite{CK06}. \emph{Unbiased die rolling} corresponds to the case $P_o=\frac{1}{N}$, for all $o\in \mathbb{Z}_N$. A die rolling protocol $\mathcal{R}$ is \emph{secure} if $\delta=0$ or if $\delta$ tends to zero by increasing some security parameter. In the former case $\mathcal{R}$ is called \emph{ideal}, while in the latter case is called \emph{arbitrarily secure}.

Blum \cite{B83} invented  \emph{coin flipping} (also called \emph{coin tossing}) in 1981, which corresponds to die rolling with $M=N=2$ and $P_0=P_1=\frac{1}{2}$, and which is  more precisely called \emph{two-party unbiased coin flipping}, and showed that it can be implemented securely with classical (non-relativistic) protocols based on  computational assumptions, like the absence of efficient protocols to factor large integers.

There exists a weak version of coin flipping and die rolling, where  each party must only be guaranteed that specific outcomes $o$ are obtained with probabilities close to $P_o$. In \emph{weak die rolling} (also called \emph{leader election} \cite{G09}) there are $M=N$ mistrustful parties at different locations rolling a $N-$faced die. For all $k\in[M]$, the $k$th party \emph{wins} if the outcome is $o=k-1$. Thus, the $k$th party must be guaranteed that if he follows the agreed protocol honestly then $\lvert P(o=k-1)-P_{k-1}\rvert\leq \delta$, for $\delta=0$ or $\delta$ tending to zero by increasing some security parameter. \emph{Weak coin flipping} corresponds to the case $M=N=2$. We note that a protocol implementing secure unbiased die rolling for $M=N$ also implements weak die rolling. But the converse is not in general true. 

To emphasize the difference between die rolling (coin flipping) and weak die rolling (weak coin flipping), the former task is sometimes called \emph{strong die rolling} (\emph{strong coin flipping}). In this paper we focus on strong die rolling and strong coin flipping, but we use the simpler terms ``die rolling'' and ``coin flipping'' to refer to these tasks.

Coin flipping and die rolling are important cryptographic task with many applications. Die rolling can be used by $M$ mistrustful parties in randomized consensus protocols, for example, to gamble, to choose a leader at random, or to fairly allocate resources in a network \cite{BD84}. It can also be used by $M$ parties to authenticate to each other remotely and securely \cite{K99.2}.


Die rolling and other cryptographic tasks are investigated in different cryptographic models, i.e. with different rules for the agreed protocols and with different constraints  on the dishonest parties,
giving rise to different security levels. The highest security level is \emph{unconditional security}, in which the dishonest parties are only constrained by the laws of physics. In particular, protocols in relativistic quantum cryptography that are provably unconditionally secure guarantee that dishonest parties who are only limited by quantum physics and the principle of no-superluminal signalling cannot break the protocols' security in close to Minkowski spacetime (like near the Earth surface) \cite{K99,K12.1}. 

\emph{No-superluminal signalling} is a fundamental physical principle of relativity theory stating that information cannot travel faster than the speed of light through vacuum in close to Minkowski spacetime. This principle is satisfied by quantum physics. In particular, two or more parties sharing an arbitrary quantum entangled state cannot communicate information faster than the speed of light by applying arbitrary quantum measurements on the quantum state.

In principle, if the dishonest parties were able to sufficiently modify the spacetime geometry then they could communicate information faster than the speed at which light travels in close to Minkowski spacetime, making the protocols in relativistic quantum cryptography insecure. However, we believe this is humanly impractical for the foreseeable future \cite{K99}. Thus, we consider that security based on quantum physics and the principle of no-superluminal signalling in approximately Minkowski spacetime is the highest level of security humanly achievable in the foreseeable future and we can then sensibly call it ``unconditional security''.

We note that relativistic quantum cryptography can in principle also be applied in spacetime geometries that are not approximately Minkowski and unconditional security can be guaranteed, if the parties know the spacetime geometry where the protocols take place with good approximation, and if there is a known upper bound on the speed of light among the protocols' locations. This requires in particular that there are no wormholes or other means allowing signalling between spacelike separated regions \cite{K99}. However, as mentioned above, we think it is sensible to assume that relativistic quantum cryptography will only be implemented by humans near the Earth surface in the foreseeable future. Thus, we think it is reasonable, and it will simplify our presentation, to assume in this paper that spacetime is approximately Minkowski.


Bit commitment \cite{M97,LC97,LC98}, oblivious transfer \cite{L97} and a class of secure two-party computations \cite{L97,BCS12} cannot achieve unconditional security with quantum non-relativistic protocols, but can be implemented securely with quantum non-relativistic protocols if the dishonest parties have bounds on the performance of their quantum memories \cite{DFSS08,WST08,NJMKW12,ENGLWW14}. Bit commitment can achieve unconditional security with classical relativistic \cite{K99,K05.2,LKBHTWZ15,CCL15,VMHBBZ16} or quantum relativistic \cite{K11.2,K12,LKBHTKGWZ13,LCCLWCLLSLZZCPZCP14,AK15.1,AK15.2} protocols. On the other hand, oblivious transfer and a class of two-party secure computations cannot achieve unconditional security even with quantum relativistic protocols \cite{CK06,C07}. However, a class of oblivious transfer protocols with constraints on the spacetime regions where the parties must obtain the outputs are provably unconditionally secure with quantum relativistic protocols \cite{K11.3,PG15.1,PGK18,PG19}.

Classical non-relativistic multi-party unbiased coin flipping protocols cannot achieve unconditional security \cite{S89,DK02}. Furthermore, they cannot unconditionally guarantee $\delta<\frac{1}{2}$ if a weak majority of the players is dishonest \cite{S89}. Thus, two-party unbiased coin flipping protocols cannot unconditionally guarantee $\delta<\frac{1}{2}$, as in this case security proofs require to assume that one party is dishonest.

Quantum non-relativistic multi-party unbiased die rolling protocols cannot achieve unconditional security either \cite{Kitaev02,ABDR04,AS10}. Furthermore, they can only unconditionally guarantee $\delta\geq \bigl(\frac{1}{N}\bigr)^{\frac{1}{M}}-\frac{1}{N}$ \cite{AS10}. This bound was first shown by Kitaev \cite{Kitaev02} for $M=N=2$ and then was generalized to $N=2$ and $M\geq 2$  by Ambainis et. al. \cite{ABDR04}, and to  $N,M\geq 2$ by Aharon and Silman \cite{AS10}. The non-existence of unconditionally secure ideal quantum-nonrelativistic protocols was shown by Lo and Chau \cite{LC98}.

Moreover, quantum non-relativistic multi-party unbiased die rolling protocols have been  shown to unconditionally guarantee $\delta= \bigl(\frac{1}{N}\bigr)^{\frac{1}{M}}-\frac{1}{N}+\epsilon$, for any even positive integer $M$ and any $N=n^\frac{M}{2}$, with any positive integer $n$ and any $\epsilon>0$ \cite{AS10}. This was first shown for two-party unbiased coin flipping ($M=N=2$) by Chailloux and Kerenidis \cite{CK09}. There exist various quantum non-relativistic protocols for two-party unbiased coin flipping that unconditionally guarantee $\delta<\frac{1}{2}$ \cite{ATVY00,SR01,NS03,A04,KN04,C07.2,CK09}. Furthermore, the optimal achievable $\delta$ that can be unconditionally guaranteed by  quantum non-relativistic multi-party unbiased coin flipping protocols with $H<M$ honest parties is $\delta=\frac{1}{2}-\Theta\bigl(\frac{H}{M}\bigr)$, i.e. satisfying $\frac{1}{2}-C_1\frac{H}{M}\leq \delta\leq \frac{1}{2}-C_2\frac{H}{M}$, for constants $C_1$ and $C_2$ with $0<C_2< C_1 < \frac{M}{2H}$ \cite{ABDR04}.


We note that, in contrast to (strong) coin flipping and (strong) die rolling, weak coin flipping \cite{M07,ACGKM16} and weak die rolling \cite{AS10} can achieve unconditional security, and arbitrarily small $\delta$, with quantum non-relativistic protocols. 

Kent \cite{K99.2} showed that unconditionally secure two-party unbiased coin flipping with arbitrarily small $\delta$ can be achieved with relativistic protocols. Colbeck and Kent \cite{CK06} introduced \emph{variable bias coin tossing}, in which one of the parties secretly chooses the bias of the coin within a stipulated range, and gave unconditionally secure quantum relativistic protocols.

Here we extend Kent's  \cite{K99.2} two-party unbiased coin flipping protocol and prove that for any integers $M,N\geq 2$ and any probability distribution $\mathcal{P}$ there exists a relativistic die rolling protocol that is unconditionally secure, with arbitrarily small $\delta$. Furthermore, we show that if the probabilities in the distribution $\mathcal{P}$ are 
rational numbers and the parties have access to perfect devices and particularly to perfectly unbiased random number generators then our protocols are ideal with unconditional security. Our results prove the claim made in Ref. \cite{CK06} -- without proof -- that all random secure two-party computations can be implemented with unconditional security. Furthermore, our results prove that this holds for an arbitrary number of parties. Our protocols do not require any quantum communication and are practical to implement with current technology. To the best of our knowledge, our protocols are the first multi-party relativistic cryptographic protocols.

\section{Security definition}

Die rolling is a task in mistrustful cryptography. In mistrustful cryptography, the parties are assumed to agree on a protocol to implement a task in collaboration, but they are not assumed to follow the agreed protocol honestly. It is in this sense that we call the parties \emph{mistrustful}. This is in contrast to quantum key distribution \cite{BB84}, for instance, where Alice and Bob collaborate with mutual trust to establish a shared key, while guaranteeing that the key remains secret to any third party.

As discussed in the introduction, in a die rolling protocol $\mathcal{R}$, $M\geq 2$ parties agree on the number $N\geq 2$ of possible outcomes $o\in\mathbb{Z}_N$, and on the ideal probability distribution $\mathcal{P}=\{P_o\}_{o=0}^{N-1}$ for the outcomes. The protocol $\mathcal{R}$ must satisfy the following two properties.

\emph{Correctness}. The protocol $\mathcal{R}$ is \emph{correct} if  all parties agree on the outcome $o$ when all parties follow $\mathcal{R}$ honestly and no party aborts.

\emph{Security}. The protocol $\mathcal{R}$ is \emph{secure} if for all $k\in[M]$, when the $k$th party follows $\mathcal{R}$ honestly and no party aborts,
then the outcome $o$ is obtained with a probability $P(o)$ satisfying  
\begin{equation}
\label{sec}
\lvert P(o)-P_o\rvert\leq \delta,
\end{equation}
for all $o\in\mathbb{Z}_N$, where $\delta=0$, or where $\delta$ tends to zero by increasing some security parameter; in the former case $\mathcal{R}$ is called \emph{ideal}, while in the latter case is called \emph{arbitrarily secure}.

An alternative figure of merit in the security definition could be the variational distance between the probability distribution $\mathcal{P}_\mathcal{R}=\{P(o)\}_{o=0}^{N-1}$ of the protocol $\mathcal{R}$ and the ideal probability distribution $\mathcal{P}$, given by
\begin{equation}
\label{vardist}
\lVert \mathcal{P}_\mathcal{R} -\mathcal{P}\rVert=\frac{1}{2}\sum_{o=0}^{N-1}\lvert P(o)-P_o\rvert.
\end{equation}
An important property of $\lVert \mathcal{P}_\mathcal{R} -\mathcal{P}\rVert$ is that the maximum probability to distinguish $\mathcal{P}_\mathcal{R}$ and $\mathcal{P}$ is given by
\begin{equation}
\label{vardist2}
P_\text{max}=\frac{1}{2}+\frac{1}{2}\lVert \mathcal{P}_\mathcal{R} -\mathcal{P}\rVert.
\end{equation}
However, we note that according to our security defintion, if $\mathcal{R}$ is secure then it holds that
\begin{equation}
\label{vardist3}
\lVert \mathcal{P}_\mathcal{R} -\mathcal{P}\rVert\leq \frac{N\delta}{2},
\end{equation}
with $\delta=0$ or with $\delta$ decreasing with some security parameter, where we used (\ref{sec}) and (\ref{vardist}).

\section{Spacetime setting}


$M$ mistrustful parties define a reference frame $F$ in near-Minkowski spacetime, for example, near the Earth surface. The parties agree in the following setting defined in $F$. We use units in which the speed of light through vacuum is unity.

Let $B_i$ be non intersecting three-dimensional balls in space with
radii $r_i$, for all $i\in[M]$. Let $d_{ij}=d_{ji}$ be the shortest distance between any point of $B_i$ and any point of $B_j$, for all $j\in[M]\setminus \{i\}$ and all $i\in[M]$. The balls are defined such that $2r_i<d_{ij}$, for all $j\in[M]\setminus\{i\}$ and all $i\in[M]$. Let $t_{i}$ be time coordinates satisfying 
\begin{equation}
\label{def3}
0<t_i<d_{ij},
\end{equation}
for all $j\in[M]\setminus\{i\}$ and all $i\in[M]$. 

For all $i,k\in[M]$, at least for the whole duration of the protocol, the $k$th party sets a secure laboratory $L_{ki}$ completely contained within $B_i$. The $k$th party does not need to trust the locations of the other parties' laboratories; but he must guarantee that his laboratory $L_{ki}$ is within $B_i$ during the protocol, for all $i,k\in[M]$.

\section{Unconditionally secure relativistic multi-party biased die rolling}


Before implementing our protocol below, the parties agree on a positive integer $n\geq N$, on $N$ non-intersecting subsets $\Omega_o$ of $\mathbb{Z}_{n}=\{0,1,\ldots,n-1\}$, for all $o\in\mathbb{Z}_N$, on a small $\alpha\geq 0$, and on small numbers $\epsilon_k\geq 0$, for all $k\in[M]$. For all $k\in[M]$, the $k$th party chooses $\epsilon_k$ such that his random number generator $R_k$ can prepare a message $m_k\in\mathbb{Z}_n$ with probability distribution $P_k(m_k)$ satisfying
\begin{equation}
\label{def2}
\biggl\lvert P_k(m_k)-\frac{1}{n}\biggr\rvert\leq \epsilon_k,
\end{equation}
for all $m_k\in\mathbb{Z}_n$, where the value of $\epsilon_k$ can correspond to the experimental uncertainty of $R_k$. The parties choose $n$ and $\Omega_o$ such that
\begin{equation}
\label{yy7}
\biggl\lvert\frac{\lvert \Omega_o\rvert}{n}-P_o\biggr\rvert\leq \alpha,
\end{equation}
for all $o\in\mathbb{Z}_N$. The parties must also guarantee that $\alpha$ and $\epsilon_k$ are sufficiently small to satisfy
\begin{equation}
\label{yy7.1}
\alpha+\epsilon_k\lvert\Omega_o\rvert\leq 1,
\end{equation}
for all $o\in\mathbb{Z}_N$ and all $k\in[M]$. From (\ref{yy7}), the parties can choose $\alpha=0$ only if $P_o$ is a rational number, for all $o\in\mathbb{Z}_N$. If not all probabilities  $P_o$ are rational numbers then the parties must choose $\alpha>0$ arbitrarily small and $n$ arbitrarily large so that (\ref{yy7}) holds. 


Our protocol comprises three stages. Stage I is a preparation stage that can take place arbitrarily in the past of stages II and III. Stage II comprises the transmission of various classical messages at spacelike separation. Finally, in stage III the parties verify that the protocol was implemented correctly and agree on the outcome $o$, after comparing the various messages received in stage II.

\begin{figure}
\includegraphics[scale=0.5]{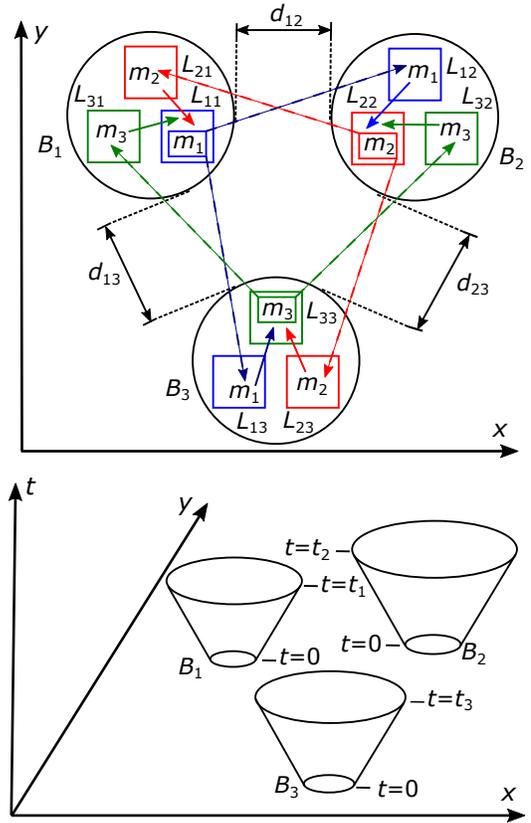}
 \caption{\label{fig1}Schematic representation of the relativistic die rolling protocol  described in the main text in $2+1$ dimensions in the reference frame $F$. The case of three parties ($M=3$) is illustrated. Up: the random number generator and classical communication channels of the first, second and third party are given in colour blue, red and green, respectively. For all $i\in[3]\setminus \{k\}$ and all $k\in[3]$, the random number generator $R_k$ outputting the message $m_k\in\mathbb{Z}_n$ is represented by the small box in the laboratory $L_{kk}$, the fast classical channel $C_{ki\rightarrow ii}$ by a short solid arrow, and the slow classical channel  $C_{kk\rightarrow ki}$ by a long dotted arrow; the channels $C_{kk\leftrightarrow ii}$ are not illustrated. The diagram is not at scale, as the balls' radii satisfy $2r_k<d_{ki}$, for all $i\in[3]\setminus \{k\}$ and all $k\in[3]$. Bottom: the balls $B_k$ at the time $t=0$ and their lightlike separated spacetime regions with time coordinates $t\in[0,t_k]$ are illustrated, for all $k\in[3]$.
}
\end{figure}

Our die rolling protocol $\mathcal{R}$ is the following (see Fig. \ref{fig1}).  

\subsection{Stage I: predistribution}
\begin{enumerate}

\item For all $k\in [M]$, $L_{kk}$ prepares a message $m_k\in\mathbb{Z}_n$ securely using a random number generator $R_k$ sufficiently before the time $t=-\max\{d_{ki}\}_{i\in[M]\setminus\{k\}}$, with probability distribution $P_k(m_k)$ satisfying (\ref{def2}), for all $m_k\in\mathbb{Z}_n$, and for some $\epsilon_k\geq 0$ satisfying (\ref{yy7}) and (\ref{yy7.1}). 

\item For all $i\in [M]\setminus\{k\}$ and all $k\in [M]$, $L_{kk}$ sends a copy of $m_k$ to $L_{ki}$ through a secure and authenticated classical channel $C_{kk\rightarrow ki}$ 
so that $L_{ki}$ receives it before the time $t=0$; the channel $C_{kk\rightarrow ki}$ does not need to be very fast because $m_k$ can be sent arbitrarily before $t=0$.

\subsection{Stage II: relativistic communication}

\item For all $i\in [M]\setminus\{k\}$ and all $k\in [M]$,  $L_{ki}$ sends $m_k$ to $L_{ii}$ through a fast classical communication channel $C_{ki\rightarrow ii}$ within the time interval $[0,t_i]$; $L_{kk}$ does not abort only if it receives $m_i\in\mathbb{Z}_n$ not after $t_k$.

\subsection{Stage III: verification}

\item For all $i\in [M]\setminus\{k\}$, all $j\in [M]\setminus\{k,i\}$ and all $k\in [M]$, $L_{kk}$ and $L_{ii}$ use a secure and authenticated classical channel $C_{kk\leftrightarrow ii}$ to verify that they received the same message $m_j\in\mathbb{Z}_n$ from $L_{jk}$ and $L_{ji}$, respectively; otherwise they abort.

\item If $L_{11},\ldots, L_{MM}$ do not abort then they agree that 
the die rolling outcome is 
\begin{equation}
\label{x6}
o=i \text{ if } x\in\Omega_i,
\end{equation}
for $i\in\mathbb{Z}_N$, where
\begin{equation}
\label{yy6}
x=\sum_{k=1}^M m_k \text{ mod } n.
\end{equation}

\end{enumerate}

We note that in the second step, the communication channel $C_{kk\rightarrow ki}$ can be implemented via secure physical transportation of the message $m_k$ from $L_{kk}$ to $L_{ki}$, or using one-time pads if $L_{kk}$ and $L_{ki}$ had been previously distributed secure keys, for all $i\in [M]\setminus\{k\}$ and all $k\in [M]$.
Because this step can be made arbitrarily in advance of the following steps, there is great flexibility on the method employed and the speed at which this is accomplished.

Similarly, the communication channels $C_{kk\leftrightarrow ii}$ between $L_{kk}$ and $L_{ii}$ in the fourth step can in principle also be implemented via secure and authenticated physical transportation of messages, for all $i\in [M]\setminus\{k\}$ and all $k\in [M]$. However, this method might not be very practical and could add undesired delays in the verification stage. For this reason, it could be preferable to implement these channels with previously distributed keys.

In principle, there could be situations where the message $m_k$ sent by the laboratory $L_{kk}$ does not reach the laboratory $L_{ki}$ at the required time, due to failure of the channel $C_{kk\rightarrow ki}$ or due to interception of the message by a dishonest party, for some $k\in[M]$ and some $i\in [M]\setminus\{k\}$. However, these situations do not arise if the channels $C_{kk\rightarrow ki}$ are secure and authenticated, for all $i\in [M]\setminus\{k\}$ and all $k\in [M]$, as we have assumed in the second step of the protocol. Nevertheless, a way to avoid these problems comprises the laboratory  $L_{ki}$ to confirm to $L_{kk}$ the reception of the message $m_k$ using a secure and authenticated channel $C_{ki\rightarrow kk}$, and the laboratory $L_{kk}$ to abort if it does not receive such a confirmation, for all $i\in [M]\setminus\{k\}$ and all $k\in [M]$. The same observations apply to the communication channels $C_{kk\leftrightarrow ii}$ in the fourth step, for all $i\in [M]\setminus\{k\}$ and all $k\in [M]$.

It is straightforward to see that the protocol is correct. If all parties follow the protocol honestly then no party aborts and all parties agree that the outcome $o$ is given by (\ref{x6}) and (\ref{yy6}).

For $k\in[M]$, below we assume that the $k$th party follows the protocol honestly 
and the other parties perform any collective quantum cheating strategy $S$, with
no party aborting.
We show that in this case the probability $P(o)$ that the die rolling outcome is $o$ satisfies
\begin{equation}
\label{yy1.1}
\lvert P(o) - P_o\rvert \leq \delta,
\end{equation}
where
\begin{equation}
\label{yyy1}
\delta=\alpha+\max_{k\in[M],o\in\mathbb{Z}_N}\{\epsilon_k\lvert\Omega_o\rvert\}.
\end{equation}
From (\ref{yy7.1}) and (\ref{yyy1}), since $\alpha\geq0$ and $\epsilon_k\geq0$, we have  $0\leq \delta\leq 1$.

Thus, if $\delta=0$, the protocol is ideal with unconditional security. From (\ref{yyy1}), this holds if $\alpha=\epsilon_k=0$ for all $k\in[M]$. From (\ref{yy7}), $\alpha=0$ can only hold if the probability distribution $\mathcal{P}=\{P_o\}_{o=0}^{N-1}$ only has rational numbers. From (\ref{def2}), $\epsilon_k=0$ can only hold if $R_k$ outputs perfectly uniform random numbers from the set $\mathbb{Z}_n$, for all $k\in[M]$.

Alternatively, if $\delta>0$ with $\delta$ decreasing exponentially by increasing some parameter then the protocol is arbitrarily secure with unconditional security. From (\ref{yyy1}), this  holds if $\alpha>0$ and $\epsilon_k>0$ decrease exponentially by increasing some security parameter, for all $k\in[M]$. From (\ref{yy7}), $\alpha$ can be chosen arbitrarily small by choosing $n$ arbitrarily large. Furthermore, from the Piling up Lemma \cite{Pilinguplemma}, if $R_k$ produces $n_\text{bits}$ bits with biases from the range $\bigl(0,\frac{1}{2}\bigr)$ then $m_k\in\mathbb{Z}_n$ can be obtained from  these bits with $\epsilon_k$ decreasing exponentially with $n_{\text{bits}}$, for all $k\in[M]$.

\subsection{The unbiased case}

It is straightforward to see that the unbiased case, i.e $P_o=\frac{1}{N}$ for all $o\in\mathbb{Z}_N$, can be implemented by setting $n=N$, $\alpha=0$ and $\Omega_o=\{o\}$, for all $o\in\mathbb{Z}_N$, in which case the die rolling outcome is $o=x$.

\subsection{Security Proof}

For all $k\in[M]$, we assume that the $k$th party follows the protocol honestly 
and the other parties implement an arbitrary collective quantum cheating strategy $S$, with no party aborting, 
and show (\ref{yy1.1}). 

We have the following properties in the frame $F$, for all $i\in[M]\setminus\{k\}$ and all $k\in[M]$.

\begin{itemize}
\item The $k$th party generates $m_k\in\mathbb{Z}_n$ securely in one of his laboratories and communicates it to his other laboratories using secure and authenticated classical channels.

\item The $k$th party sends $m_k$ to the $i$th party at a spacetime region $Q_{ki}$ with spatial coordinates in $B_i$ and time coordinates from the interval $[0,t_i]$.

\item The $k$th party does not abort in step 3 only if he receives a message $m_i\in\mathbb{Z}_n$ within a spacetime region $Q_{ik}$ with spatial coordinates in $B_k$ and time coordinates not greater than $t_k$.

\item From (\ref{def3}), the shortest distance $d_{ki}$ between any point in $B_k$ and any point in $B_i$ satisfies $0<t_k<d_{ki}$. Thus, the spacetime regions  $Q_{ki}$ and $Q_{ik}$ are spacelike separated (see Fig. \ref{fig1}).
\end{itemize}

The principle of no-superluminal signalling states that information cannot travel faster than the speed of light through vacuum in close to Minkowski spacetime. Thus, from the previous points,
the probability to obtain the set of messages $\tilde{m}_k=\{m_i\vert i\in[M]\setminus\{k\}\}$ given the message $m_k$, is given by
\begin{equation}
\label{x0}
P_k^S(\tilde{m}_k\vert m_k)=P_k^S(\tilde{m}_k),
\end{equation}
for all $m=(m_1,m_2,\ldots,m_M)\in\mathbb{Z}_n^M$ and all $k\in[M]$. That is, the probability distribution for $\tilde{m}_k$ is independent of $m_k$. We note that this holds for an arbitrary quantum cheating strategy by the dishonest parties. In particular, in a general quantum cheating strategy, the laboratories of all the parties (honest and dishonest) may share an arbitrary entangled quantum state $\lvert\psi\rangle$, and each laboratory may apply an arbitrary quantum measurement on its share of $\lvert\psi\rangle$ in order to obtain its input or trying to communicate received information to laboratories at other locations. However, the principle of no-superluminal signalling, and consequently (\ref{x0}), hold in this general situation.

It follows from (\ref{x0}) that the probability distribution for the string of messages $m$
satisfies
\begin{equation}
\label{x0.1}
P_{kS}(m)=P_k^S(\tilde{m}_k\vert m_k)P_{k}(m_k)=P_k^S(\tilde{m}_k)P_{k}(m_k),
\end{equation}
for all $m\in\mathbb{Z}_n^M$ and all $k\in[M]$, were in the second equality we used (\ref{x0}).
We define
\begin{equation}
\label{yy3.1}
\Delta_k(y)\equiv\biggl\{\tilde{m}_k\in\mathbb{Z}_n^{M-1}\bigg\vert \sum_{i\neq k} m_i=y \text{ mod } n\biggr\},
\end{equation} 
for all $k\in[M]$ and all $y\in\mathbb{Z}_n$.

From  (\ref{x6}), (\ref{yy6}), (\ref{x0.1}) and (\ref{yy3.1}), the probability $P(o)$ that the die rolling outcome is $o$ satisfies
\begin{equation}
\label{yy4}
P(o)=\sum_{x\in\Omega_o}\sum_{y\in\mathbb{Z}_n}\sum_{\tilde{m}_k\in\Delta_k(x-y)}P_k(m_k=y \text{ mod } n)P_{k}^S(\tilde{m}_k),
\end{equation}
for all $o\in\mathbb{Z}_N$. In (\ref{yy4}), we sum over all strings $\tilde{m}_k$ satisfying that the sum of their entries $m_i$ equals $x-y \text{~mod~}n$; we also sum over all possible values $y\text{~mod~}n$ for $m_k$, and over all $x\in\Omega_o$. From (\ref{yy4}), we have
\begin{eqnarray}
\label{yy4.2}
P(o)&\leq& \biggl(\frac{1}{n}+\epsilon_k\biggr) ~\sum_{x\in\Omega_o}\sum_{y\in\mathbb{Z}_n}~ \sum_{\tilde{m}_k\in\Delta_k(x-y)}P_{k}^S(\tilde{m}_k)\nonumber\\
&=& \biggl(\frac{1}{n}+\epsilon_k\biggr) \sum_{x\in\Omega_o}~\sum_{\tilde{m}_k\in\mathbb{Z}_n^{M-1}} P_{k}^S(\tilde{m}_k)\nonumber\\
&=& \biggl(\frac{1}{n}+\epsilon_k\biggr) \lvert\Omega_o\rvert\nonumber\\
&\leq&P_o+\alpha+\epsilon_k\lvert\Omega_o\rvert\nonumber\\
&\leq&P_o+\delta,
\end{eqnarray}
for all $o\in\mathbb{Z}_N$, where in the first line we used (\ref{def2}); in the second line we used (\ref{yy3.1}); in the third line we used that 
\begin{equation}
\label{neweq}
\sum_{\tilde{m}_k\in\mathbb{Z}_n^{M-1}}P_{k}^S(\tilde{m}_k)=1;
\end{equation}
in the fourth line we used (\ref{yy7}); and in the last line we used (\ref{yyy1}). Similarly, it follows straightforwardly that
\begin{equation}
\label{yy5}
P(o)\geq P_o-\delta,
\end{equation}
for all $o\in\mathbb{Z}_N$. Thus, (\ref{yy1.1}) follows from (\ref{yy4}) and (\ref{yy5}).

\subsection{Composability}

An important security property in cryptography is that of composable security. Broadly speaking, for a cryptographic protocol to have composable security, not only must it be secure when implemented on its own, but it must also be composed as a secure subroutine for more general cryptographic tasks. 

According to Ref. \cite{VPR19}, coin flipping (and bit commitment) cannot achieve composable security even with the most general type of protocols in relativistic quantum cryptography. The argument of Ref. \cite{VPR19} is that a condition for a coin flipping protocol to have composable security is that the outcome $o$ of the protocol must be independent of the outcome $o'$ of another arbitrary coin flipping protocol that may take place in parallel, and that this condition cannot be guaranteed because a dishonest party may apply a man-in-the-middle attack and correlate the outcomes $o$ and $o'$ of both protocols. We discuss below how this argument applies to our die rolling protocols.

Consider the case of two parties ($M=2$). Alice and Bob implement a die rolling protocol $\mathcal{R}$ giving outcome $o$. Bob and Charlie implement another die rolling protocol $\mathcal{R}'$ in parallel, with the same parameters of $\mathcal{R}$, giving outcome $o'$. Let Alice and Charlie be honest and let Bob be dishonest. Bob implements the following man-in-the-middle attack with the effect that $o=o'$ \cite{VPR19,PR21}. 

The protocols $\mathcal{R}$ and $\mathcal{R}'$ are implemented in the space balls $B_1$ and $B_2$ in parallel with arbitrarily small time delays. Bob plays the role of the second party in $\mathcal{R}$ and the role of the first party in $\mathcal{R}'$, whereas Alice  plays the role of the first party in $\mathcal{R}$, and Charlie plays the role of the second party in $\mathcal{R}'$. In $B_2$, Alice sends the message $m_1$ to Bob in the protocol $\mathcal{R}$; Bob then sends the message $m_1'=m_1$ to Charlie in the protocol $\mathcal{R}'$. In $B_1$, Charlie sends the message $m_2'$ to Bob in the protocol $\mathcal{R}'$; Bob then sends the message $m_2=m_2'$ to Alice in the protocol $\mathcal{R}$. Thus, from (\ref{yy6}), we obtain that the value of $x$ in $\mathcal{R}$ and its corresponding value $x'$ in $\mathcal{R}'$, given by
\begin{eqnarray}
\label{q1}
x &=& m_1+m_2 \text{ mod } n,\nonumber\\
\label{q2}
x' &=& m_1'+m_2' \text{ mod } n, 
\end{eqnarray}
satisfy $x'=x$. Thus, from (\ref{x6}), we have $o=o'$. 

However, the previous man-in-the-middle attack does not apply if Alice and Charlie both play the roles of the first (second) party and they are guaranteed to be honest. Alternatively, we may consider Alice and Charlie to be the same party, playing the role of the first (second) party in both $\mathcal{R}$ and $\mathcal{R}'$ and performing the protocols honestly. In this case, if Alice and Charlie play the roles of say the first party then their messages $m_1$ and $m_1'$ given to Bob in $B_2$ are independent. Thus, because Bob must give messages $m_2$ and $m_2'$ to Alice and Charlie in $B_1$ at spacelike separation from Alice and Charlie giving Bob the messages $m_1$ and $m_1'$ in $B_2$, it follows that the messages $m_2$ and $m_2'$ are independent of $m_1$ and $m_1'$. Thus, from (\ref{yy6}) and (\ref{q1}), $x$ and $x'$ are independent. It follows from (\ref{x6}) that the die rolling outcomes $o$ and $o'$ are also independent.

The previous arguments can be extended straightforwardly to the case of $M>2$ parties. In this case, if we assume that the $k$th party is honest in the protocol $\mathcal{R}$ and the $j$th party is honest in the protocol $\mathcal{R}'$, with $k\neq j$, and all other parties are dishonest and collaborate as a single party in both protocols $\mathcal{R}$ and $\mathcal{R}'$, then a simple extension of the previous man-in-the-middle attack implies that the die rolling outcomes of $\mathcal{R}$ and $\mathcal{R}'$ satisfy $o=o'$.

However, if the $k$th party is honest in both protocols $\mathcal{R}$ and $\mathcal{R}'$ then the previous attack does not apply, and the $k$th party can be guaranteed that the outcomes $o$ and $o'$ are independent. This is because in this case, in the protocol $\mathcal{R}$ ($\mathcal{R}'$), the $k$th party gives the $i$th party a message $m_k$ ($m_k'$) in the space ball $B_i$ at spacelike separation from the $k$th party receiving a message $m_i$ ($m_i'$) from the $i$th party in the space ball $B_k$, for all $i\in[M]\setminus\{k\}$. Thus, the messages $m_i$ and $m_i'$ are independent of $m_k$ and $m_k'$, for all $i\in[M]\setminus\{k\}$. Furthermore, since the $k$th party is honest in $\mathcal{R}$ and $\mathcal{R}'$, the messages $m_k$  and $m_k'$ are independent. It follows from (\ref{yy6}) that the respective values of $x$ and $x'$ in $\mathcal{R}$ and $\mathcal{R}'$ are independent, which implies from (\ref{x6}) that the respective outcomes $o$ and $o'$ of $\mathcal{R}$ and $\mathcal{R}'$ are also independent.

It follows from Ref. \cite{VPR19} and from the previous discussion that our die rolling protocols cannot be composed securely in arbitrary ways. However, a honest party can participate in various die rolling protocols in parallel and be guaranteed not only that each protocol is secure on its own, but also that the outcomes of the protocols are independent, by choosing carefully the spacetime regions where she communicates her messages to, and where she accepts messages from, the other parties in all the protocols where she is participating. We have discussed this possibility above for die rolling protocols taking place in spacetime regions that are arbitrarily close when the honest party plays the role of the $k$th party in all protocols, for some $k\in[M]$. Another straightforward way to guarantee to a honest party that the outcomes of different die rolling protocols $\mathcal{R}$ and $\mathcal{R}'$ are independent is that every spacetime region where she receives or communicates a message in $\mathcal{R}$ is spacelike separated from every spacetime region where she receives or communicates a message in $\mathcal{R}'$.
Given these observations, we believe that whether our die rolling protocols can be composed securely to implement other cryptographic tasks, perhaps under some assumptions and/or in relativistic settings, deserves further investigation.

\section{Discussion}


Apart from achieving unconditional security with arbitrarily small $\delta$, our protocols have the following advantages over existing quantum non-relativistic coin flipping and die rolling protocols.

First, our protocols do not require any quantum communication. Thus, they are free from various experimental challenges of quantum non-relativistic coin flipping and die rolling protocols, like noise and losses \cite{BM04,BM04.2,TVUZ05,LBABM05,NFHM08,BBBG09,C10,AMS10,BBBGST11,PJLCLTKD14}, and side-channel and multi-photon attacks \cite{BCDKPG21}.

Second, the distance among the parties can be made very large in practice in our protocols, with laboratories spread on the Earth surface or in satellites orbiting the Earth, for instance. For example, Ref. \cite{LKBHTKGWZ13} experimentally demonstrated a relativistic protocol with laboratories separated by $9354$ km. This cannot be easily achieved with protocols involving quantum communication. For example, Refs. \cite{BBBGST11,PJLCLTKD14} experimentally demonstrated two-party quantum non-relativistic coin flipping protocols achieving a security advantage over classical non-relativistic protocols at distance separations of only a few meters and 15 km, respectively.

As already mentioned, weak coin flipping and weak die rolling can achieve unconditional security and arbitrarily small $\delta$ with quantum non-relativistic protocols \cite{M07,ACGKM16,AS10}. Our protocols trivially implement these tasks with unconditional security too, and have the advantages mentioned above.

Nevertheless, implementing our protocols have some important challenges. First, some communication steps must be very fast to achieve spacelike separation. However, this is feasible with field programmable gate arrays if the protocol sites are sufficiently far apart, as demonstrated by Refs. \cite{LKBHTKGWZ13,LCCLWCLLSLZZCPZCP14,LKBHTWZ15,VMHBBZ16,ABCDHSZ21}. 
In particular Ref. \cite{ABCDHSZ21} experimentally demonstrated a relativistic cryptographic protocol with locations separated by only $60$ m.

Second, the laboratories must be synchronized securely to a common reference frame with sufficient time precision. This can be achieved with GPS devices and atomic clocks \cite{LKBHTKGWZ13,LCCLWCLLSLZZCPZCP14,LKBHTWZ15,VMHBBZ16,ABCDHSZ21}, for instance.

We note that a dishonest party can in principle implement attacks in the reference frame synchronization of the other parties, for example, by spoofing their GPS signals. If these attacks are implemented successfully without being caught, the parties under attack may believe that the communications in the relativistic stage were implemented at spacelike separation, while in fact they were not, in this way compromising the protocol's security. To our knowledge, previous experimental demonstrations of relativistic cryptography have been vulnerable to these attacks \cite{LKBHTKGWZ13,LCCLWCLLSLZZCPZCP14,LKBHTWZ15,VMHBBZ16,ABCDHSZ21}. A countermeasure against these attacks is for each party to synchronize her clocks in a secure laboratory and then distribute them securely to her other laboratories, guaranteeing that the clocks remain sufficiently synchronized during the relativistic stage of the protocol \cite{VMHBBZ16}.

Our protocols are intrinsically classical but can be made quantum by using quantum random number generators. Ideally, the parties use quantum random number generators to guarantee that their inputs are truly random. Additionally, quantum key distribution links \cite{BB84,FLDCTPSYS17,Liaoetal17} can be used to expand the secure keys shared among the various laboratories used to implement the long distance communication channels.
This can be suitably implemented in quantum networks \cite{ECPPSY05,S17,Liaoetal18,DWTSTLPYDCTEGWP19} or a quantum internet \cite{K08,WER18}.

\begin{acknowledgments}
The author acknowledges financial support from the UK Quantum Communications Hub grant no. EP/T001011/1.

\end{acknowledgments}


%

\end{document}